\documentclass[aps,prl,showpacs,twocolumn,nofootnoteinbib,groupedaddress]{revtex4}
\usepackage{epsfig,amssymb,amsmath,latexsym}
\newcommand\sone{${{S_1}}$~}

\newcommand\stwo{${{S_2}}$~}

\newcommand\dthree{${{D_3}}$~}

\newcommand\dfourd{${{D_4}}$}
\newcommand\dfour{${{D_4}}$~}

\newcommand\prightd{${{P_R}}$}
\newcommand\pright{${{P_R}}$~}
\newcommand\pleftd{${{P_L}}$}
\newcommand\pleft{${{P_L}}$~}

\newcommand\pfour{${{P_4}}$~}

\newcommand\pthree{${{P_3}}$~}

\newcommand\aoneal{${{a_1^\alpha}}$~}

\newcommand\atwoal{${{a_2^\alpha}}$~}

\newcommand\aonebe{${{a_1^\beta}}$~}

\newcommand\athreeal{${{a_3^\alpha}}$~}

\newcommand\afourald{${{a_4^\alpha}}$}

\newcommand\atwobe{${{a_2^\beta}}$~}

\newcommand\aleftal{${{a_L^\alpha}}$~}

\newcommand\arightal{${{a_R^\alpha}}$~}

\newcommand\pleftalbe{${{P}}_{ L}^ {\alpha \beta}$~}

\newcommand\prightalbe{${{P}}_{ R}^ {\alpha \beta}$~}

\def\ket#1{| \,#1\, \rangle}

\def\expect#1{\langle \,#1\, \rangle}
\def\scx#1#2{\langle \,#1\, |\, #2\, \rangle}

\def\dfrac#1#2{{\displaystyle\frac{#1}{#2}}}

\def\lsim{\mathrel{\rlap{\lower4pt\hbox{\hskip1pt$\sim$}}
    \raise1pt\hbox{$<$}}}         
\def\gsim{\mathrel{\rlap{\lower4pt\hbox{\hskip1pt$\sim$}}
    \raise1pt\hbox{$>$}}}         

\newcommand{\beq}{\begin{equation}}
\newcommand{\eeq}{\end{equation}}
\newcommand{\bea}{\begin{eqnarray}}
\newcommand{\eea}{\end{eqnarray}}

\def\lsim{\:\raisebox{-0.5ex}{$\stackrel{\textstyle<}{\sim}$}\:}
\def\gsim{\:\raisebox{-0.5ex}{$\stackrel{\textstyle>}{\sim}$}\:}

\newcommand\hbtd{{\sf{HB-T}}}
\newcommand\hbt{{\sf{HB-T}~}}

\newcommand\poi{Poincar$\acute{\rm e}$~}

\begin{document}
\textheight=23.8cm
\title{\Large
 The Nonlocal Pancharatnam Phase in Two-Photon Interferometry}
\author{\bf Poonam Mehta}
\email{poonam@rri.res.in}
\author{\bf Joseph Samuel}
\email{sam@rri.res.in}
\author{\bf Supurna Sinha}
\email{supurna@rri.res.in}
 \affiliation{Raman Research Institute, Bangalore 560 080, India}
\date{\today}
\pacs{03.65.Vf, 
 25.75.Gz,   
 95.75.Kk   
 42.50.-p 
 }
%
\begin{abstract}
We propose a polarised intensity interferometry experiment, which
measures the nonlocal Pancharatnam phase acquired by a {\it pair}
of Hanbury Brown-Twiss photons. The setup involves two
polarised thermal sources illuminating two polarised detectors.
Varying the relative polarisation angle of the detectors introduces
a two photon geometric phase. Local measurements at either
detector do not reveal the effects of the phase, which is an optical
analog of the multiparticle Aharonov-Bohm effect.
The geometric phase sheds light on the three slit
experiment and suggests ways of tuning entanglement.
\end{abstract}
%
\maketitle

{{\emph{Introduction :-}}} The familiar two slit experiment in
quantum mechanics describes the interference of a single particle
with itself. However, there are also quantum processes that describe
the interference of a pair of particles with itself. As shown by
Hanbury Brown and Twiss (\hbtd)~\cite{hbt} about fifty years ago,
such interference is observed in the coincidence counts of photons.
Their original motivation was to measure the diameters of stars
replacing Michelson interferometry by intensity interferometry.
Their work was initially met with scepticism because the quantum
mechanical interpretation of the proposed experiment was unclear at
the time. The resulting controversy led to the birth of the new
field of quantum optics. Intensity interferometry is now routinely
used in a variety of fields, from nuclear physics~\cite{baym} to
condensed matter~\cite{heiblum}.

In the nineteen eighties, Berry discovered~\cite{berry} the
geometric phase in quantum mechanics, which has now been applied and
studied in various contexts~\cite{shaperebook}. It was soon realised
that Berry's discovery had been anticipated by Pancharatnam's
work~\cite{panch} on the interference of polarised light,
Pancharatnam's work is now widely recognised as an early precursor
of the geometric phase~\cite{ramaseshan}, with a perspective that is
far more general~\cite{sam} than the context in which it was
discovered by Berry.

B{\"u}ttiker~\cite{buttiker1} noted in the context of electronic
charge transport that two-particle correlations can be sensitive to
a magnetic flux even if the single particle observables are flux
insensitive. The effect of the flux is visible only in current cross
correlations and is a genuinely nonlocal and multiparticle
Aharonov-Bohm effect~\cite{ab}. This effect has been experimentally
seen in intensity interferometry experiments carried out using edge
currents in quantum Hall systems~\cite{heiblum} and the theory was
further developed in~\cite{samuelsson,buttiker3} and the possibility
of controlled orbital entanglement and the connection to Bell
inequalities mentioned.

In  this paper, we  propose a new experiment with polarised light,
which shows geometric phase effects {\it only in the intensity
correlations} ${\cal G}^2$ and not in the lower order correlations
${\cal G}^1$. The two photon Pancharatnam phase effect is also
nonlocal in the precise sense that it cannot be seen by local
measurements at either detector. Coincidence detection of photons in
two detectors yields counts which are modulated by a phase which has
a geometric component as well as the expected dynamical (or
propagation) phase. Unlike in earlier
studies~\cite{brendel,hariharanbook}, the effects of the geometric
phase are seen {\it only} in the {\it cross} correlation counts of
two detectors. Neither the count rate nor self correlation of the
individual detectors shows any such geometric phase effects. The
phase is given by half the solid angle enclosed on the \poi sphere
by the total circuit of a {\it pair} of \hbt photons and as
expected, is {\it achromatic}.

 The experimental setup is described below and then a theoretical analysis
  is given. Finally we conclude with a discussion and a comparison with previous work.

\begin{figure}[htb]
\centering \vspace*{3mm} \hspace*{1mm} \epsfig{figure=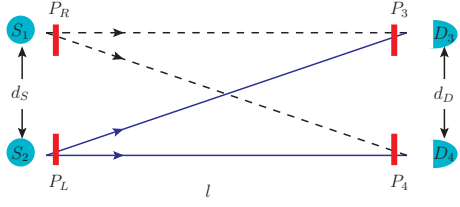,width
=0.76\columnwidth} \vspace*{-1mm} \caption{Schematic diagram of the
proposed experiment :
 \sone and \stwo are thermal sources, covered by circular analysers
which only pass right and left circular light respectively. The two
detectors \dthree and \dfour receive only linear polarizations. The
angle $\varphi_{34}$ between the axes of the linear polarizers can
be continuously varied.  The dashed and solid lines represent
photons from the two sources \sone and \stwo respectively. The
separation between the detectors is $d_D$ and that between the
sources is $d_S$. } \label{hbtfig1}
\end{figure}

{{\emph{Proposed Experiment :-}}} The experiment consists of having
two thermal sources \sone and \stwo illuminate two detectors \dthree
and \dfour (Fig.~\ref{hbtfig1}). This setup is very similar to the
\hbt experiment~\cite{hbt}. The only difference is in the use of
analysers (shown in red online), which select a particular state of
polarisation. The source \sone is covered by an analyser \prightd,
which only permits Right Hand Circular  light to pass through it,
while the source \stwo is covered by an analyser \pleftd, which only
permits Left Hand Circular light to pass through. The light is
incident on detectors \dthree and \dfour after passing through
polaroids \pthree and \pfour respectively that only permit linearly
polarised light to pass through (linear analysers). The angle
${\varphi}_{34}$ between
the axes of \pthree and \pfour and the detector separation $d_D$ can
be continuously varied in the experiment. The measured quantity is
the coincidence count ${\cal C}$ of photons received at detectors
\dthree and \dfourd, \bea && {\cal C} ={\cal G}^2=
\dfrac{\expect{N_3 N_4}}{\expect{N_3}\expect{N_4}}~, \label{corr}
\eea
where $N_3$ and $N_4$ are the photon numbers detected at \dthree and
\dfour per unit time (per unit bandwidth).

\begin{figure}[htb]
\centering \vspace*{3mm} \hspace*{1mm} \epsfig{figure=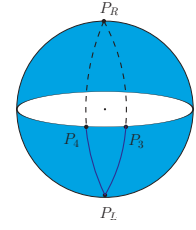,width
=0.3\columnwidth} \vspace*{-1mm} \caption{The path on the \poi
sphere that determines the geometric phase. The angle $\varphi_{34}$
between the linear polaroids determines the width of the lune on the
\poi sphere and the geometric phase.} \label{hbtfig2}
\end{figure}

As in the \hbt interferometer, we would expect the coincidence
counts to vary with the propagation phases and so the counts would
depend on the detector separation $d_D$ and the wavelength $\lambda$
of the light. The new effect that is present in the polarised
version is that we expect the coincidence counts to also depend on
$\varphi_{34}$  and to be modulated by a geometric phase of half the
solid angle on the \poi sphere shown in Fig.~\ref{hbtfig2}.

The geometric phase is achromatic, unlike the propagation phases
mentioned above. Note that the path traversed on the \poi sphere is
not traced by a {\it single} photon, but by  a {\it pair} of HBT
photons. Thus the experiment explores a new {\it avatar} of the
geometric phase in the context of intensity interferometry.


{{\emph{Theory :-}}} We now calculate the expected coincidence
counts for the detectors \dthree and \dfour and show that these
counts depend on the geometric phase. For ease of calculation we
suppose that we are dealing with a single frequency {\it{i.e.}} a
quasi-monochromatic beam. In fact the detectors will have a finite
acceptance bandwidth, which has to be incorporated in a more
realistic calculation. The principle of the effect comes across
better in the present idealised situation.

We write \aoneal and \atwoal for the destruction operators of the
photon modes at the sources \sone and \stwo where $\alpha$ runs over
the two polarisation states. The modes just after the analysers
\pright and \pleft are represented by projections \arightal =
\prightalbe \aonebe and
 \aleftal = \pleftalbe \atwobe where a sum over repeated Greek
indices is understood and the projection matrices \pright and \pleft
onto the right and left circular states represent the action of the
analysers. The modes at the detectors are characterised by the
destruction operators \athreeal and \afourald.
%
%
We suppose that the separation $l$ between the sources and the
detectors is much larger than the separation  $d_S$ between the
sources and the separation $d_D$ between the detectors {\it i.e.},
$l >> d_S, d_D$. When light is emitted by a source and received by a
detector, it suffers propagation  phases and decrease of its
amplitude inversely with distance. These effects are captured in the
functions $u_{ij}=\frac{1}{l} \exp\{i[k(|{\vec r}_i-{\vec
r}_j|)-\omega t]\}$, where $\omega$ is the frequency of the light,
$k$ is the wave vector and ${\vec r}_i$ and ${\vec r}_j$ the
locations of the detector and source.
 With this
notation, we express $a_{b}^\alpha$ (where $b=3,4$) as $
{{a_b^\alpha}} = {{P}}_b^{\alpha\beta}~[\,{{P}}_{ L}^{\beta\gamma}
\, {{a_2^{\gamma} }}\, u_{b2} \,+ \, {{P}}_{R}^{\beta\gamma} \,
{{a_1^{\gamma}}}\, u_{b1}\,]$ and its Hermitean conjugate
$a_b^{\dagger \alpha}$ as ${{a_b^{\dagger \alpha}}} = [\,{\bar
u}_{b2}\, {{a_2^{\dagger \gamma}}}\, {{P}}_{ L}^{\gamma \beta} \,+\,
{\bar u}_{b1}\, {{a_1^{\dagger\gamma}}}\, {{P}}_{R}^{\gamma\beta}\,]
~ {{P}}_b^{\beta\alpha}$ where the overbar stands for complex
conjugation and we use the fact that the $2\times2$ Hermitean
projection matrices ${{P}}$ satisfy ${{P}}^2={P}$ and ${\bar
{{P}}^{\alpha\beta}}=P^{\beta\alpha}$.

The quantities of interest~\footnote{For any general operator $\hat
{\cal O}$, $\expect{\hat {\cal{O}}}$ = ${\rm{Tr}}[\varrho \, \hat
{\cal{O}}]$ where $\varrho$ is the normalised thermal density matrix
$\varrho=\exp\{-\beta H \}/Z$ with $Z = {\rm{Tr}} [\exp\{-\beta
H\}]$ and $H=(a^\dagger a+1/2) \omega $.} are $\expect{N_3}$,
$\expect{N_4}$, being the photon counts per unit time (per unit
bandwidth) at the two detectors ($D_3$ and $D_4$) and $\expect{:N_3
N_4:}$, the coincidence counts, where the $:\, :$ stands for normal
ordering which has to be applied to the number operator product in
the numerator of Eq.~(\ref{corr}). $N_b$ is given by $N_b =
a_b^{\dagger \alpha} a_b^\alpha =
 \big[\,
\bar{u}_{b2}\;a^{\dagger\alpha}_{2}\;(P_{L}P_{b})^{\alpha\beta} +
\bar{u}_{b1}\;a^{\dagger\alpha}_{1}\;(P_{R}P_{b})^{\alpha\beta}
\,\big]
 \big[\, P^{\beta \gamma}_{L}\;a^{\gamma}_{2} u_{b2} +
P^{\beta\gamma}_{R}\;a^{\gamma}_{1} u_{b1} \,\big]$.
We  find
 \bea \expect{N_{b}} &=&
\bar{u}_{b1}\;u_{b1}\;(P_{R} P_{b}P_{R})^{\alpha\beta}\;
\expect{a^{\dagger\alpha}_{1} a_{1}^{\beta}} \nonumber\\ && +
\bar{u}_{b2}\;u_{b2}\; (P_{L} P_{b}P_{L})^{\alpha\beta}\;
\expect{a^{\dagger\alpha}_{2} a_{2}^{\beta}}~. \label{N3}\eea From
the thermal nature of the
sources, 
$\expect{a^{\dagger\alpha}_{1} a_{1}^{\beta} }
=\expect{a^{\dagger\alpha}_{2} a_{2}^{\beta} } =
\delta^{\alpha\beta} \, n_B$
 where $ n_B$ is the Bose function
${( \exp\{\beta\hbar\omega\} - 1 ) }^{-1}$ and $\beta$ the inverse
temperature. And we arrive at $\expect{N_{3}} = \expect{N_{4}} =
{n_B}/{l^{2}}$. The computation of $\expect{: N_3 N_4 :}$ is
slightly more involved but straightforward. The product $N_{3}
N_{4}$  is a product of four brackets each of which has two terms.
When the four brackets are expanded, there are sixteen terms, of
which ten vanish. The six nonzero terms combine to give
\begin{widetext}\bea \expect{:N_3 N_4:} = n_B^{2} \left[
\frac{3}{2\,l^{4}} + \bar{u}_{32}\; u_{31}\; \bar{u}_{41}\; u_{42}\;
{\rm Tr}\left[P_{L}P_{3}P_{R}P_{4}P_{L}\right] + \bar{u}_{31}\;
u_{32}\; \bar{u}_{42}\; u_{41}\; {\rm
Tr}\left[P_{R}P_{3}P_{L}P_{4}P_{R}\right] \right]~. \label{n3n4}
\eea
\end{widetext} Only the second and third terms in Eq.~(\ref{n3n4})
contain the propagation and geometric phases. The sequence of
projections can be viewed as a series of closed loop quantum
collapses~\cite{ramaseshan,sam} given by $\scx{R}{3} \scx{3}{L}
\scx{L}{4}
\scx{4}{R}$
\bea {\rm Tr} \left[P_{R}P_{3}P_{L}P_{4}P_{R}\right] &=&
\frac{1}{4} \exp{\left\{ i
\frac{\Omega}{2} \right\}}~, \label{collapses} \eea
 where $\Omega$ is the solid angle subtended by the
geodesic path $\ket{R} \to \ket{3} \to \ket{L} \to \ket{4} \to
\ket{R} $ at the center of the \poi sphere. Apart from the phase,
the projections also result in an amplitude factor of
$1/4$~\cite{panch} since projections are non-unitary operations
leading to a loss in intensity.  The final theoretical expression
for ${\cal C}$
in the limit $l
>> d_S,d_D$ is \bea {\cal C} &=& \frac{3}{2} +
\frac{1}{2} \cos\left[\vec d_D \cdot (\vec k_2 - \vec k_1) +
\frac{\Omega}{2}\right]~,\label{final}
 \eea  where $\vec k_i = k \hat r_i$ is the
wavevector of light seen in the $i^{\rm {th}}$ detector. (The
propagation phases in Eq.~(\ref{final}) can also be written in an
equivalent form with the sources and detectors exchanged.).  It is
also easily seen that the self correlation
 $\expect{: N_3 N_3 :}$ ($\expect{: N_4 N_4  :}$) can be obtained by
  replacing $4$ by $3$  ($3$ by $4$)  in Eq.~(\ref{n3n4})
 above. In this case, the sequence of projections $
 {\rm Tr} \left[P_{R}P_{3}P_{L}P_{3}P_{R}\right]$ ($
 {\rm Tr} \left[P_{R}P_{4}P_{L}P_{4}P_{R}\right]$)
 subtends a zero solid angle and the geometric contribution to the phase
 vanishes.
 Thus neither the photon counts $\expect{N_3}$, $\expect{N_4}$ in
 individual detectors nor the self correlations $\expect{: N_3 N_3   :}$,
 $\expect{: N_4 N_4  :}$  reveal the
geometric phase. This supports our claim that the effect described
here is only present in the cross-correlations and  {\it{not}} in
the self correlations.

${\cal C}$ depends on the experimentally tunable parameters $d_D$
and $\varphi_{34}$. The geometric part is achromatic and depends
only on $\varphi_{34}$. The propagation part in the phase carries
the dependence on $d_D$ as well as on the wavelength. By changing the
angle $\varphi_{34}$ between the axes of the two polaroids, we can
conveniently modulate the geometric component $\Omega$. If the
propagation and geometric phases are set to zero, we find that the
correlation ${\cal C}$ takes the value $2$, just as in original \hbt
interferometry.


{{\emph{Conclusion :-}}} We have proposed a simple generalisation of
the \hbt experiment which uses the vector nature of light to produce
a geometric phase. The only difference between the proposed
experiment and the \hbt experiment is the presence of polarisers at
the sources and detectors. These polarisers cause a geometric phase
to appear in the coincidence counts of two detectors which receive
linearly polarised light. Neither the count rates nor the self
correlations of individual detectors show any geometric phase
effects. These appear solely in the {\it cross} correlations in the
count rates of the detectors. The appearance of the geometric phase
cannot be attributed or localised to any single segment joining a
source ($S_1,S_2$) to a detector $(D_3,D_4)$. It appears only when
one considers the {\it two photon path} (Fig.~2) on the \poi sphere
in its entirety. Our experiment brings out a new result of a
conceptual nature, which may not have been guessed without our
present understanding of the Pancharatnam phase. The experiment
proposed here would be a good demonstration of a {\it purely
multiparticle and nonlocal} geometric phase in optics. We hope to
interest experimentalists in this endeavour. Apart from verifying
the theoretical expectation, our proposed experiment suggests
further lines of thought concerning multiparticle and nonlocal
effects which may be stimulating to research in this area. We
mention two of these, the first an application of our ideas to
generating controlled entanglement and the second of a more
conceptual nature regarding the role of probabilities in quantum
mechanics.

{\it Controlled Entanglement :} The experimental setup described
above can be used to make a source of photon pairs with a controlled
degree of entanglement. Like many other elementary particles, the
photon has spin (polarisation) as well as orbital (spacetime)
degrees of freedom. Our idea is to use the polarisation degree of
freedom to control the orbital entanglement of photons. Let us
replace the two thermal sources of Fig.~\ref{hbtfig1} by a single
two-photon source producing a pair of oppositely circularly
polarised  photons. Each photon is then passed through an
interferometric delay line which consists of a short and a long arm
with time delays $t_S$ and $t_L$. The relative amplitudes and phases
of the two paths can be chosen to generate any state in the two
dimensional Hilbert space spanned by $\ket{S}$ and $\ket{L}$. By
such means we can arrange for the incident state at $P_R$ to be in a
spin state of right circular polarisation and in an orbital state
$\ket{\phi}_1 = \alpha \ket{S}_1 + \beta\ket{L}_1$ and similarly,
the incident  state at $P_L$ to be in a spin state of left circular
polarisation and in an orbital state $\ket{\psi}_2 = \alpha'
\ket{S}_2 + \beta'\ket{L}_2$, where $\alpha,\beta$ {\it{etc}} are
complex numbers. The input state  is therefore a direct product of
states at $P_R$ and $P_L$: $\ket{\phi}_1\otimes\ket{\psi}_2$. By
combining the amplitudes for the two photons to arrive at the
detectors via the paths $1-3,2-4$ and $1-4,2-3$ (direct and
exchange) we find that the state at the output is of the form
$\ket{\phi}_3\otimes \ket{\psi}_4 + \exp\{i \Omega/2\}
~\ket{\psi}_3\otimes\ket{\phi}_4$, where the geometric phase factor
$\exp\{i\Omega/2\}$ is the relative phase between the direct and
exchange processes. This final two photon state is entangled as it
cannot in general be written as a direct product $\ket{\Psi}_3
\otimes \ket{\Phi}_4$ of photon states at $3$ and $4$. The
entanglement is generated by particle exchange effects rather than
interactions. The degree of entanglement can be tuned using the
polaroid setting $\varphi_{34}$. The degree of entanglement can be
quantified either using Bell's inequality or by the von Neumann
entropy of the reduced density matrix after tracing over one of the
subsystems ($3$ or $4$). A straightforward calculation of the von
Neumann entropy shows that it does depend on the geometric phase.
Since the geometric phase is achromatic, we can apply the same phase
over all the frequencies in the band of interest by tuning
$\varphi_{34}$ and generate entangled photon pairs with a degree of
precision and control. This setup can be used as a source of
entangled photon pairs for other experiments probing quantum
entanglement.

{\it The Three-Slit Experiment :}
Quantum mechanics is often introduced by a discussion of the two-slit
experiment in which an electron is incident on an opaque barrier with
two slits and then detected when it falls on a screen.
The surprise of the quantum theory is that
the outcome of the two-slit experiment is not determined by the outcome
of one-slit experiments in which one or the other of the slits is
blocked. This is
in sharp contrast to classical random processes like Brownian motion. If
one considers the passage of a Brownian particle through slits $A$ and $B$,
one finds that~\footnote{Throughout this discussion we neglect
single particle trajectories that recross the barrier and wind around
multiple slits.}
\begin{equation}
{\cal P}_{AB}={\cal P}_A+{\cal P}_B~,\nonumber
\end{equation}
where ${\cal P}_{AB}$ is the probability
of detecting the particle when both slits are open and ${\cal P}_A$ and
${\cal P}_B$ are the corresponding detection probabilities in one-slit
experiments. Thus classical probabilities are one slit separable, but quantum
probabilities are not: the equality above is not satisfied
in the two slit quantum experiment.

However, if one considers three slits $A,B,C$, one finds that in quantum
mechanics, the outcome of the three-slit experiment {\it is} determined
by the outcomes of the one and two slit experiments. Mathematically,
\begin{equation}
{\cal P}_{ABC}={\cal P}_{AB}+{\cal P}_{BC}+{\cal P}_{CA}-{\cal P}_A-{\cal P}_B-{\cal P}_C~,\nonumber \end{equation}
which  follows easily from writing ${\cal P}_{ABC}=|\psi_A+\psi_B+\psi_C|^2$
and $\psi_A,\psi_B,\psi_C$ are the amplitudes for passage through
the slits. Thus quantum mechanics is two slit separable~\cite{sorkin-1994-9}.
This is why we do not find a discussion of the three slit experiment
in elementary Quantum Mechanics books: it brings in nothing new.

The situation changes when one considers multiparticle
and nonlocal processes of the kind
exemplified by our experiment of Fig.~\ref{hbtfig1}. Consider
a {\it three}-slit experiment in which three incoherent beams
of light fall upon three slits $A,B,C$ which are covered by
analysers $P_A,P_B,P_C$ each of which allows a single state on
the \poi sphere to pass. The light from the analysers is then
allowed to fall on three unpolarised detectors labelled $4,5,6$.
By considerations similar to our analysis of the experiment of
Fig.~\ref{hbtfig1}, we find that the number correlations
$\expect{N_4 N_5 N_6}$ contain terms involving the geometric
phase (half the solid angle subtended by the three polarisation
states $A,B,C$ of the analysers). Such an effect is not present in
any of the two-slit or one-slit experiments, since two (or fewer)
points on the \poi sphere do not enclose  a solid angle. The effect
is a genuinely three slit effect, not decomposable in terms of two
and one slit effects. Thus quantum theory contains effects which
are not two slit separable because of multiparticle entanglement.
Our three slit experiment involving the geometric phase brings
out this point forcefully.

The question of whether a single particle crossing a barrier with
slits obeys two slit separability is ultimately an experimental one.
The theoretical possibility of violations of two slit separability
in such experiments was noted by Sorkin~\cite{sorkin-1994-9}, who
proposed that there may be theories going beyond quantum mechanics
which admit such effects. There have been
attempts~\cite{urbasisinha} to search for such effects in a three
slit experiment using photons. Since these experiments are null
experiments, one has to be careful to rule out all possible three
slit effects that are present due to multiparticle entanglement.
Geometric phase effects which involve three photons are an example
of such three slit effects. The experiment we propose here in
Fig.~\ref{hbtfig1} is just the simplest of a class of phenomena
involving multiparticle entanglement, nonlocality and the geometric
phase. We hope to interest the quantum optics community in pursuing
these ideas further.

\acknowledgments{It is a pleasure to thank Urbasi Sinha for
discussions related to the three-slit experiment, Anders Kastberg
for discussions on a possible experimental realisation and Hema
Ramachandran and R. Srikanth for discussions on entanglement.}

\bibliographystyle{apsrev}
\bibliography{referenceshbt}

\begin{thebibliography}{16}
\expandafter\ifx\csname natexlab\endcsname\relax\def\natexlab#1{#1}\fi
\expandafter\ifx\csname bibnamefont\endcsname\relax
  \def\bibnamefont#1{#1}\fi
\expandafter\ifx\csname bibfnamefont\endcsname\relax
  \def\bibfnamefont#1{#1}\fi
\expandafter\ifx\csname citenamefont\endcsname\relax
  \def\citenamefont#1{#1}\fi
\expandafter\ifx\csname url\endcsname\relax
  \def\url#1{\texttt{#1}}\fi
\expandafter\ifx\csname urlprefix\endcsname\relax\def\urlprefix{URL }\fi
\providecommand{\bibinfo}[2]{#2}
\providecommand{\eprint}[2][]{\url{#2}}

\bibitem[{\citenamefont{Hanbury~Brown and Twiss}(1956)}]{hbt}
\bibinfo{author}{\bibfnamefont{R.}~\bibnamefont{Hanbury~Brown}}
  \bibnamefont{and} \bibinfo{author}{\bibfnamefont{R.~Q.} \bibnamefont{Twiss}},
  \bibinfo{journal}{Nature} \textbf{\bibinfo{volume}{177}}, \bibinfo{pages}{27}
  (\bibinfo{year}{1956}).

\bibitem[{\citenamefont{Baym}(1998)}]{baym}
\bibinfo{author}{\bibfnamefont{G.}~\bibnamefont{Baym}}, \bibinfo{journal}{Acta.
  Phys. Polonica} \textbf{\bibinfo{volume}{B29}}, \bibinfo{pages}{1839}
  (\bibinfo{year}{1998}).

\bibitem[{\citenamefont{Neder et~al.}(2007)\citenamefont{Neder, Ofek, Chung,
  Heiblum, Mahalu, and Umansky}}]{heiblum}
\bibinfo{author}{\bibfnamefont{I.}~\bibnamefont{Neder}},
  \bibinfo{author}{\bibfnamefont{N.}~\bibnamefont{Ofek}},
  \bibinfo{author}{\bibfnamefont{Y.}~\bibnamefont{Chung}},
  \bibinfo{author}{\bibfnamefont{M.}~\bibnamefont{Heiblum}},
  \bibinfo{author}{\bibfnamefont{D.}~\bibnamefont{Mahalu}}, \bibnamefont{and}
  \bibinfo{author}{\bibfnamefont{V.}~\bibnamefont{Umansky}},
  \bibinfo{journal}{Nature} \textbf{\bibinfo{volume}{448}},
  \bibinfo{pages}{333} (\bibinfo{year}{2007}).

\bibitem[{\citenamefont{Berry}(1984)}]{berry}
\bibinfo{author}{\bibfnamefont{M.~V.} \bibnamefont{Berry}},
  \bibinfo{journal}{Proc. Roy. Soc. Lond.} \textbf{\bibinfo{volume}{A392}},
  \bibinfo{pages}{45} (\bibinfo{year}{1984}).

\bibitem[{\citenamefont{Shapere and Wilczek}(1989)}]{shaperebook}
\bibinfo{author}{\bibfnamefont{A.}~\bibnamefont{Shapere}} \bibnamefont{and}
  \bibinfo{author}{\bibfnamefont{F.}~\bibnamefont{Wilczek}},
  \emph{\bibinfo{title}{Geometric Phases in Physics}}
  (\bibinfo{publisher}{World Scientific, Singapore}, \bibinfo{year}{1989}).

\bibitem[{\citenamefont{Pancharatnam}(1956)}]{panch}
\bibinfo{author}{\bibfnamefont{S.}~\bibnamefont{Pancharatnam}},
  \bibinfo{journal}{Proc. Ind. Acad. Sci.} \textbf{\bibinfo{volume}{A44}},
  \bibinfo{pages}{247} (\bibinfo{year}{1956}).

\bibitem[{\citenamefont{Ramaseshan and Nityananda}(1986)}]{ramaseshan}
\bibinfo{author}{\bibfnamefont{S.}~\bibnamefont{Ramaseshan}} \bibnamefont{and}
  \bibinfo{author}{\bibfnamefont{R.}~\bibnamefont{Nityananda}},
  \bibinfo{journal}{Curr. Sci.} \textbf{\bibinfo{volume}{55}},
  \bibinfo{pages}{1225} (\bibinfo{year}{1986}).

\bibitem[{\citenamefont{Samuel and Bhandari}(1988)}]{sam}
\bibinfo{author}{\bibfnamefont{J.}~\bibnamefont{Samuel}} \bibnamefont{and}
  \bibinfo{author}{\bibfnamefont{R.}~\bibnamefont{Bhandari}},
  \bibinfo{journal}{Phys. Rev. Lett.} \textbf{\bibinfo{volume}{60}},
  \bibinfo{pages}{2339} (\bibinfo{year}{1988}).

\bibitem[{\citenamefont{B\"uttiker}(1992)}]{buttiker1}
\bibinfo{author}{\bibfnamefont{M.}~\bibnamefont{B\"uttiker}},
  \bibinfo{journal}{Phys. Rev. Lett.} \textbf{\bibinfo{volume}{68}},
  \bibinfo{pages}{843} (\bibinfo{year}{1992}).

\bibitem[{\citenamefont{Aharonov and Bohm}(1959)}]{ab}
\bibinfo{author}{\bibfnamefont{Y.}~\bibnamefont{Aharonov}} \bibnamefont{and}
  \bibinfo{author}{\bibfnamefont{D.}~\bibnamefont{Bohm}},
  \bibinfo{journal}{Phys. Rev.} \textbf{\bibinfo{volume}{115}},
  \bibinfo{pages}{485} (\bibinfo{year}{1959}).

\bibitem[{\citenamefont{Samuelsson et~al.}(2004)\citenamefont{Samuelsson,
  Sukhorukov, and B\"uttiker}}]{samuelsson}
\bibinfo{author}{\bibfnamefont{P.}~\bibnamefont{Samuelsson}},
  \bibinfo{author}{\bibfnamefont{E.~V.} \bibnamefont{Sukhorukov}},
  \bibnamefont{and}
  \bibinfo{author}{\bibfnamefont{M.}~\bibnamefont{B\"uttiker}},
  \bibinfo{journal}{Phys. Rev. Lett.} \textbf{\bibinfo{volume}{92}},
  \bibinfo{pages}{026805} (\bibinfo{year}{2004}).

\bibitem[{\citenamefont{Splettstoesser
  et~al.}(2009)\citenamefont{Splettstoesser, Moskalets, and
  B\"uttiker}}]{buttiker3}
\bibinfo{author}{\bibfnamefont{J.}~\bibnamefont{Splettstoesser}},
  \bibinfo{author}{\bibfnamefont{M.}~\bibnamefont{Moskalets}},
  \bibnamefont{and}
  \bibinfo{author}{\bibfnamefont{M.}~\bibnamefont{B\"uttiker}},
  \bibinfo{journal}{Phys. Rev. Lett.} \textbf{\bibinfo{volume}{103}},
  \bibinfo{pages}{076804} (\bibinfo{year}{2009}).

\bibitem[{\citenamefont{Brendel et~al.}(1995)\citenamefont{Brendel, Dultz, and
  Martienssen}}]{brendel}
\bibinfo{author}{\bibfnamefont{J.}~\bibnamefont{Brendel}},
  \bibinfo{author}{\bibfnamefont{W.}~\bibnamefont{Dultz}}, \bibnamefont{and}
  \bibinfo{author}{\bibfnamefont{W.}~\bibnamefont{Martienssen}},
  \bibinfo{journal}{Phys. Rev. A} \textbf{\bibinfo{volume}{52}},
  \bibinfo{pages}{2551} (\bibinfo{year}{1995}).

\bibitem[{\citenamefont{Hariharan}(2003)}]{hariharanbook}
\bibinfo{author}{\bibfnamefont{P.}~\bibnamefont{Hariharan}},
  \emph{\bibinfo{title}{Optical Interferometry}} (\bibinfo{publisher}{Elsevier
  Science (U.S.A.)}, \bibinfo{year}{2003}).

\bibitem[{\citenamefont{Sorkin}(1994)}]{sorkin-1994-9}
\bibinfo{author}{\bibfnamefont{R.~D.} \bibnamefont{Sorkin}},
  \bibinfo{journal}{Mod. Phys. Lett. A} \textbf{\bibinfo{volume}{9}},
  \bibinfo{pages}{3119} (\bibinfo{year}{1994}).

\bibitem[{\citenamefont{Sinha et~al.}(2010)\citenamefont{Sinha, Couteau,
  Jennewein, Laflamme, and Weihs}}]{urbasisinha}
\bibinfo{author}{\bibfnamefont{U.}~\bibnamefont{Sinha}},
  \bibinfo{author}{\bibfnamefont{C.}~\bibnamefont{Couteau}},
  \bibinfo{author}{\bibfnamefont{T.}~\bibnamefont{Jennewein}},
  \bibinfo{author}{\bibfnamefont{R.}~\bibnamefont{Laflamme}}, \bibnamefont{and}
  \bibinfo{author}{\bibfnamefont{G.}~\bibnamefont{Weihs}},
  \bibinfo{journal}{Science} \textbf{\bibinfo{volume}{329}},
  \bibinfo{pages}{418} (\bibinfo{year}{2010}).

\end{thebibliography}

\end{document}